\documentclass[sigconf]{acmart}

\usepackage{paralist}
\usepackage{listings}
\usepackage{circledsteps}


\usepackage{tikz}
\usetikzlibrary{shapes,arrows}
\usetikzlibrary{calc}
\usetikzlibrary{fit,positioning}
\usetikzlibrary{patterns}

\AtBeginDocument{%
  }

\copyrightyear{2023}
\acmYear{2023}
\setcopyright{licensedothergov}\acmConference[ICPE '23 Companion]{Companion of the 2023 ACM/SPEC International Conference on Performance Engineering}{April 15--19, 2023}{Coimbra, Portugal}
\acmBooktitle{Companion of the 2023 ACM/SPEC International Conference on Performance Engineering (ICPE '23 Companion), April 15--19, 2023, Coimbra, Portugal}
\acmPrice{15.00}
\acmDOI{10.1145/3578245.3584851}
\acmISBN{979-8-4007-0072-9/23/04}

\begin{document}

\title{Towards Solving the Challenge of Minimal Overhead Monitoring}

\author{David Georg Reichelt}
\email{d.g.reichelt@lancaster.ac.uk}
\orcid{0000-0002-1772-1416}
\affiliation{%
  \institution{Lancaster University in Leipzig / Universität Leipzig}
  \streetaddress{P.O. Box 1212}
  \city{Leipzig}
  \state{Saxony}
  \country{Germany}
  \postcode{04105}
}

\author{Stefan Kühne}
\email{stefan.kuehne@uni-leipzig.de}
\orcid{0000-0001-9492-2548}
\affiliation{%
  \institution{Universität Leipzig}
  \streetaddress{P.O. Box 1212}
  \city{Leipzig}
  \state{Saxony}
  \country{Germany}
  \postcode{04105}
}

\author{Wilhelm Hasselbring}
\email{hasselbring@email.uni-kiel.de}
\orcid{0000-0001-6625-4335}
\affiliation{%
  \institution{Kiel University}
  \streetaddress{1 Th{\o}rv{\"a}ld Circle}
  \city{Kiel}
  \state{Schleswig-Holstein}
  \country{Germany}}

\renewcommand{\shortauthors}{David Georg Reichelt, Stefan Kühne, \& Wilhelm Hasselbring}

\begin{abstract}
  The examination of performance changes or the performance behavior of a software requires the measurement of the performance. This is done via probes, i.e., pieces of code which obtain and process measurement data, and which are inserted into the examined application. The execution of those probes in a singular method creates overhead, which deteriorates performance measurements of calling methods and slows down the measurement process. Therefore, an important challenge for performance measurement is the reduction of the measurement overhead.
  
  To address this challenge, the overhead should be minimized. Based on an analysis of the sources of performance overhead, we derive the following four optimization options: 
\begin{inparaenum}
  \item Source instrumentation instead of AspectJ instrumentation,
  \item reduction of measurement data,
  \item change of the queue and
  \item aggregation of measurement data.
\end{inparaenum}
We evaluate the effect of these optimization options using the MooBench benchmark. Thereby, we show that these optimizations options reduce the monitoring overhead of the monitoring framework Kieker. For MooBench, the execution duration could be reduced from 4.77 $\mu s$ to 0.39 $\mu s$ per method invocation on average.
\end{abstract}

\ccsdesc[500]{General and reference~Performance}
\ccsdesc[500]{Software and its engineering~Software performance}
\ccsdesc[300]{Software and its engineering~Software maintenance tools}

\keywords{software performance engineering, benchmarking, performance measurement, monitoring overhead}


\maketitle

\section{Introduction}
The examination of performance changes, performance anomalies or the general performance behavior of a software usually requires the measurement of the performance.\footnote{Alternative approaches require data sources with similarly detailed resolution, e.g. using data from existing logging \cite{liao2022locating} requires fine-grained log statements.} The measurement in production environments is called monitoring \cite{waller2015performance}.\footnote{Nowadays, the detailed measurement in production environments is often also called \textit{observability}, since tool marketing wants to emphasize the comprehensiveness of obtained data and possible analysis. We stick to the classical term \textit{monitoring}.} Performance monitoring is possible using two options:
\begin{inparaenum}
  \item using instrumentation, i.e., by inserting measurement probes into the source code, or
  \item using sampling, i.e., by frequently obtaining the current state of the application from the outside.
\end{inparaenum} 
Measurement probes are pieces of code which obtain and process measurement data, and which are inserted into the examined application. Frequently obtaining the current state of the application requires a suitable technical interface to do so, e.g. the JVM provides the tool \lstinline'jstack', which obtains the current stack of an application.

Both, instrumentation and sampling, create overhead. When measuring the performance, this overhead reduces the accuracy of the measurements itself and slows down the measurement process. If the accuracy of measurements decreases, this is often resulting in increased standard deviation of the measurement values. The detectability of small performance changes depends on the effect size, i.e., the relation between performance change size and standard deviation. Therefore, especially detection of small performance changes require minimal performance monitoring overhead. This is especially important when aiming for the identification of small performance changes \cite{reichelt2019asedemo}. Hence, an important \textit{challenge} for performance monitoring is the reduction of the measurement overhead. 
  
To address this challenge, the main options are:
\begin{inparaenum}
  \item the optimization of sampling intervals and configuration,
  \item the configuration of adaptive monitoring supported by adaptive instrumentation and
  \item the optimization of the measurement process itself.
\end{inparaenum}
In this work, we present a first step addressing the challenge of minimal monitoring overhead by reducing the overhead of the measurement process itself (option~3). We demonstrate these overhead reductions using the application performance monitoring framework Kieker \cite{Kieker2020}. 

Based on an analysis of the sources of performance overhead in Kieker, we derive the following four optimizations:
\begin{inparaenum}[(1)]
  \item Source code instrumentation, i.e., instrumenting the source code of the software instead of using AspectJ,
  \item Reduced data storage, i.e., only measuring the duration and the method name,
  \item Queue exchange, i.e., use a \lstinline'CircularFifoQueue' instead a \lstinline'LinkedBlockingQueue', and
  \item Aggregated writing, i.e., writing the aggregated measured performance data instead writing the measurement values of every singular method call.
\end{inparaenum}
We apply the empirical standard of benchmarking \cite{hasselbring2021benchmarking} to compare the effect of these overhead reductions, and find that each optimization is able to significantly reduce the performance overhead. Using all overhead reductions, we can reduce the overhead from 4.77 $\mu s$ to $0.39 \mu s$ per method invocation.

The remainder of this paper is organized as follows: First, we give an introduction of the Kieker framework and the MooBench benchmark. Afterwards, we describe our optimizations for the Kieker probes. Subsequently, we provide measurement results of the overheads using MooBench. Our results are then compared to related work. Finally, we give a summary and outlook.

\section{Foundations}

In this section, we describe the monitoring framework Kieker and the monitoring overhead benchmark MooBench.

\subsection{Kieker}

A variety of tools is able to perform performance monitoring, including Kieker, SPASSmeter \cite{eichelberger2014flexible}, inspectIT,\footnote{\url{https://www.inspectit.rocks/}} OpenTelemetry\footnote{\url{https://opentelemetry.io/}} and Dynatrace APM.\footnote{\url{https://www.dynatrace.com/de/platform/application-performance-monitoring/}} The OpenAPM\footnote{\url{https://openapm.io/}} initiative gives an overview over existing monitoring tools and their interoperability. 

Two central components of a monitoring tool are the \textit{library} and the \textit{agent}. One of both or both are added to the execution of a software in order to obtain monitoring data. These monitoring data might be at infrastructure, application or business level; in this work, we focus on \textit{application} monitoring. Application monitoring can be done either using instrumentation, i.e., adding monitoring code to the measured system, or using sampling, i.e., obtaining measurement data at given times by technical interfaces.

Sampling in the JVM is only able to obtain monitoring data at safepoints \cite{hofer2015lightweight}. Therefore, it stops the execution of all threads and might obtain inaccurate measurement data. To avoid these pitfalls, we focus on monitoring using instrumentation. According to the MooBench benchmark, Kieker is the framework with the lowest monitoring overhead when compared to OpenTelemetry and inspectIT \cite{reichelt2021ssp}.

\begin{figure}
    \begin{tikzpicture}[node distance = 1em, auto, font=\footnotesize, align=center, every node/.style={line width=1pt,draw,shape=rectangle,minimum width = 2.75cm, align=center}]

      \node (Original) [shape=rectangle]{Original \\source};
      \node (Probe) [shape=rectangle, below=0cm of Original]{\Circled{2} Probe};
      \node (Instrumentierung) [shape=rectangle, left=1cm of Original]{\Circled{1} Instrumentation};
      \node (Queue) [shape=rectangle, below=of Probe]{\Circled{3} Queue};
      \node (Writer) [shape=rectangle, below=of Queue]{\Circled{4} Writer};
      \node (Festplatte) [shape=rectangle, below=0.5cm of Writer]{Hard Disk};
      
        \node (InstrumentierungOpt) [shape=rectangle, above=0.5cm of Instrumentierung, fill=green]{AspectJ};      
        \draw[thick] (InstrumentierungOpt) -- (Instrumentierung);
        \node (ProbeOpt) [shape=rectangle, above=0.5cm of Original, fill=green]{OperationExecutionRecord};      
        \draw[thick] (ProbeOpt.east) edge [bend left=30] (Probe.east);
        \node (QueueOpt) [shape=rectangle, left=1cm of Queue, fill=green]{LinkedBlockingQueue};      
        \draw[thick] (QueueOpt) -- (Queue);
        \node (WriterOpt) [shape=rectangle, left=1cm of Writer, fill=green]{FileWriter};      
        \draw[thick] (WriterOpt) -- (Writer);
      
      \draw[draw, rounded corners, line width=2pt] ($(Original.north west) - (4.25, -1.5cm)$) rectangle ($(Original.north east) - (-0.5, 3.25cm)$) node [draw=none,pos=.10,xshift=-1.7cm] {\textbf{JVM}};

      \draw [->] (Instrumentierung) -- (Original);
      \draw [->] (Instrumentierung) -- (Probe.west);

      \draw [->] (Probe) -- (Queue);
      \draw [->] (Queue) -- (Writer);

      \draw [->] (Writer) -- (Festplatte);
    \end{tikzpicture}
    \caption{Monitoring Process in Kieker}
    \label{fig:monitoringAblauf}
\end{figure}
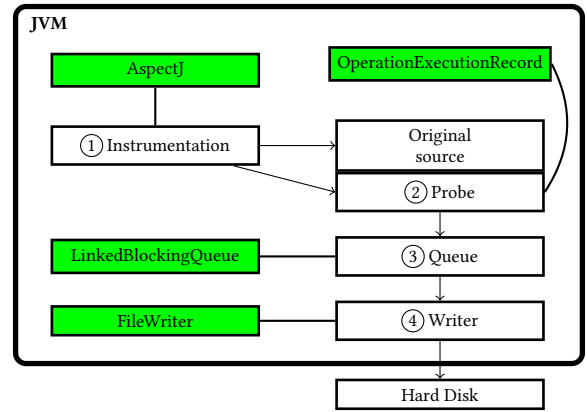

The measurement process of Kieker is depicted in Figure~\ref{fig:monitoringAblauf}. At first, the instrumentation is executed, which is regularly done using AspectJ. This adds a monitoring probe, which creates the monitoring records, e.g., one instance of \lstinline'OperationExecutionRecord' for every method invocation. The records contain metadata of the execution, e.g., execution order index and execution stack size, which enable reconstruction of the call tree afterwards. These records are passed to the queue afterwards, which is a \lstinline'LinkedBlockingQueue' by default. In parallel to the program main thread, a writer is executed in a parallel thread. The writer thread awaits new records and writes the records to their destination. By default, this is a \lstinline'FileWriter' writing the monitoring records to the hard disk.

\subsection{MooBench}

\begin{figure}
  \begin{tikzpicture}[node distance = 1em, auto, font=\footnotesize, align=center, every node/.style={line width=1pt,draw,shape=rectangle,minimum width = 1.0cm}]
    \node (main) [shape=circle]{main};
    \node (c0) [shape=circle, right=of main, inner sep=1pt]{$method$\\$(t, d)$};
    \node (c1) [shape=circle, right=of c0, inner sep=1pt]{$method$\\$(t, d-1)$};
    \node (cp) [shape=circle, right=of c1]{...};
    \node (cn) [shape=circle, right=of cp]{$method$\\$(t,1)$};
    \node (work) [shape=circle, right=of cn]{$method$\\$(t,0)$};  
 
    \draw [->] (main) -- (c0);
    \draw [->] (c0) -- (c1);
    \draw [->] (c1) -- (cp);
    \draw [->] (cp) -- (cn);
    \draw [->] (cn) -- (work);
  \end{tikzpicture}
  \caption{Call Tree of MooBench}
  \label{fig:MooBench}
\end{figure}
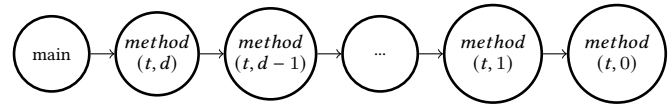

MooBench is a benchmark that aims for measuring the performance overhead of monitoring frameworks \cite{Waller2015}. To measure the performance overhead, it calls a method recursively with a given call depth $d$. In the leaf node, it executes busy waiting until a specified amount of time $t$ has passed. To avoid performance optimizations, the recursive method calls return the timestamp of the root method call. The call tree of this process is visualized in Figure~\ref{fig:MooBench}.\footnote{Original source code: \url{https://github.com/kieker-monitoring/moobench/blob/ef9ca00259a8546e6fa9cdda47ac4cca3f116fe5/tools/benchmark/src/main/java/moobench/application/MonitoredClassSimple.java}}

Performance measurement in Java is affected by non-determinism due to Just-in-Time Compilations, optimizations and garbage collections of the JVM. Therefore, a complex measurement process and a statistical analysis of measurements is required \cite{georges2007statistically}. 

Moobench therefore contains scripts which automate the repetition of VM starts and benchmark iterations inside the VM, so performance differences can be detected in a statistically sound manner. Originally, it was built to monitor Kieker, inspectIT and SpassMETER \cite{eichelberger2014flexible}. For each of the supported frameworks, MooBench contains a definition of different measurement configurations, e.g., measuring the overhead of performance measurement in Kieker with serialization to hard disk or to a TCP receiver. This makes it possible to compare the performance of these variants. MooBench has been used in different execution environments \cite{knoche2018using}. Recently, it has been extended for measurement of OpenTelemetry \cite{reichelt2021ssp}.

\section{Performance Optimizations for Kieker Probes}

\begin{figure}
    \begin{tikzpicture}[node distance = 1em, auto, font=\footnotesize, align=center, every node/.style={line width=1pt,draw,shape=rectangle,minimum width = 2.75cm, align=center}]

      \node (Original) [shape=rectangle]{Original \\source};
      \node (Probe) [shape=rectangle, below=0cm of Original]{\Circled{2} Probe};
      \node (Instrumentierung) [shape=rectangle, left=1cm of Original]{\Circled{1} Instrumentation};
      \node (Queue) [shape=rectangle, below=of Probe]{\Circled{3} Queue};
      \node (Writer) [shape=rectangle, below=of Queue]{\Circled{4} Writer};
      \node (Festplatte) [shape=rectangle, below=0.5cm of Writer]{Hard Disk};
      
        \node (InstrumentierungOpt) [shape=rectangle, above=0.5cm of Instrumentierung, fill=green]{Source Code\\ Instrumentation};      
        \draw[thick] (InstrumentierungOpt) -- (Instrumentierung);
        \node (ProbeOpt) [shape=rectangle, above=0.5cm of Original, fill=green]{DurationRecord};      
        \draw[thick] (ProbeOpt.east) edge [bend left=30] (Probe.east);
        \node (QueueOpt) [shape=rectangle, left=1cm of Queue, fill=green]{CircularFifoQueue};      
        \draw[thick] (QueueOpt) -- (Queue);
        \node (WriterOpt) [shape=rectangle, left=1cm of Writer, fill=green]{AggregatedWriter};      
        \draw[thick] (WriterOpt) -- (Writer);
      
      \draw[draw, rounded corners, line width=2pt] ($(Original.north west) - (4.25, -1.5cm)$) rectangle ($(Original.north east) - (-0.5, 3.25cm)$) node [draw=none,pos=.10,xshift=-1.7cm] {\textbf{JVM}};

      \draw [->] (Instrumentierung) -- (Original);
      \draw [->] (Instrumentierung) -- (Probe.west);

      \draw [->] (Probe) -- (Queue);
      \draw [->] (Queue) -- (Writer);

      \draw [->] (Writer) -- (Festplatte);
    \end{tikzpicture}
    \caption{Possible Monitoring Optimizations}
    \label{fig:monitoringOptimierungen}
\end{figure}
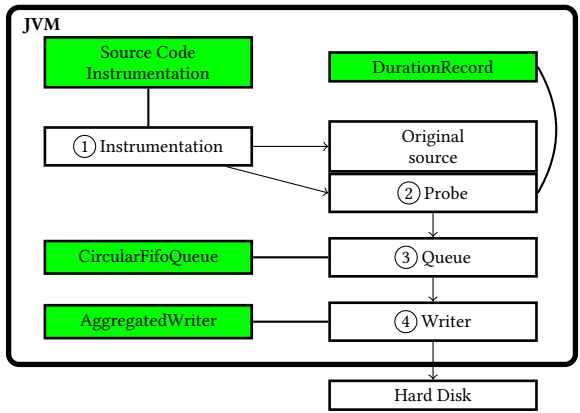

All the parts of Kieker's monitoring process provide room for performance optimization:
\begin{inparaenum}
  \item To avoid the AspectJ overhead, \textbf{source code instrumentation} can be used,
  \item to avoid metadata creation steps, only the necessary data to create a \textbf{DurationRecord} can be obtained,
  \item to decrease waiting times for queue inserts, a \textbf{CircularFifoQueue} can be used and
  \item to decrease waiting times for data storage, only \textbf{aggregated data} can be stored.
\end{inparaenum}
A summary of the optimizations is visualized in Figure~\ref{fig:monitoringOptimierungen}. The details of these optimizations are described in the following subsections.

\subsection{Source Code Instrumentation Instead AspectJ}

Monitoring the program execution, e.g., measuring the execution time of methods, requires changes to the executed source. This can be done using instrumentation libraries, like ByteBuddy for OpenTelemetry\footnote{\url{https://github.com/open-telemetry/opentelemetry-java-instrumentation/blob/main/javaagent-extension-api/build.gradle.kts}} or AspectJ in Kieker.\footnote{\url{https://kieker-monitoring.readthedocs.io/en/1.15.2/getting-started/AspectJ-Instrumentation-Example.html\#gt-aspectj-instrumentation-example}} Kiekers AspectJ instrumentation is configured via the \lstinline'aop.xml'. It defines which aspect should be used, i.e., whether \lstinline'OperationExecutionRecord's, capturing start and end time of a method execution, or \lstinline'BeforeOperationRecords', capturing only the start of a method cexecution, is used. When starting the application, the monitoring is started by passing the parameter \lstinline'-javaagent:kieker-1.15.2-aspectj.jar' to the JVM.

Using instrumentation libraries creates overhead, because it creates \lstinline'JoinPointImpl.proceed' calls to the stack trace. The overhead can be avoided by directly inserting the monitoring code into the monitored application. Therefore, we created the tool \textit{kieker-source-instrumentation}\footnote{\url{https://github.com/kieker-monitoring/kieker-source-instrumentation}} which automatically adds measurement code to all called methods. 
This requires two automated changes to the monitored application's source code:
\begin{inparaenum}
  \item Adding the required variables to the monitored class and
  \item adding the monitoring source code to all methods that should be monitored.
\end{inparaenum}

The first step is necessary, since monitoring requires at least a reference to the \lstinline'MonitoringController', to pass the created monitoring record to the queue, and the currently used \lstinline'TimeSource', to get the current time. These are both singletons; however, obtaining the instances on the fly using \lstinline'getInstance' creates overhead and should be avoided. Therefore, the source instrumentation creates \lstinline'static final' fields for every class that is instrumented.

For the second step, the monitoring source code is inserted into the method. This always requires definition of the signature, determining start and end time of the method, creating a monitoring record and writing this record to the queue. Optionally, other metadata are obtained, like the execution stack size. All variables are created with a prefix, e.g. \lstinline'_kieker', so collisions with existing method names are avoided. A simplified example of instrumentation with determining the current stack size is depicted in Listing~\ref{lst:instrumentation}.

\begin{lstlisting}[caption=Example Instrumented Source, label=lst:instrumentation,float=*]
public void myMethod(){
  final String _kieker_signature = "public void net.kieker.Class.myMethod()";
  [...]
  final int _kieker_ess;
  long _kieker_traceId = _kieker_controlFlowRegistry.recallThreadLocalTraceId();
  [...]
  if (_kieker_traceId == -1) {
    _kieker_ess = _kieker_controlFlowRegistry.recallAndIncrementThreadLocalESS();
    [...]
  } else {
    _kieker_ess = _kieker_controlFlowRegistry.recallAndIncrementThreadLocalESS();
    [...]
  }
  final long _kieker_tin = C0_0._kieker_TIME_SOURCE.getTime();
  try {
    // Execute method original code
  } finally {
    final long _kieker_tout = C0_0._kieker_TIME_SOURCE.getTime();
    _kieker_controller.newMonitoringRecord(
      new OperationExecutionRecord(_kieker_signature, _kieker_tin, _kieker_tout, ...));
}
\end{lstlisting}

\subsection{DurationRecord Instead of OperationExecutionRecord}

For monitoring using the \lstinline'OperationExecutionRecord', the execution order index, the execution stack size, the current hostname and an id of the current session are stored. This requires storing the current count of executions and the current stack size. Additionally, the data are stored in the record and are written to the hard disk.

Therefore, we created the \lstinline'DurationRecord', which is a minimal Kieker record that only stores the duration. To enable monitoring this record, we adapted the instrumentation process accordingly.

\subsection{CircularFifoQueue Instead of LinkedBlockingQueue}

Currently, Kieker uses a \lstinline'LinkedBlockingQueue'. It is a singly linked list that stores a capacity. If there are more elements in the queue than the capacity, the queue blocks on every new insert. Blocking slows down the execution of the application thread significantly, and should therefore never happen. A downside of the \lstinline'LinkedBlockingQueue' is, that it needs to create new queue elements, which requires reserving space, and link the new element by setting the pointer of the last queue element.

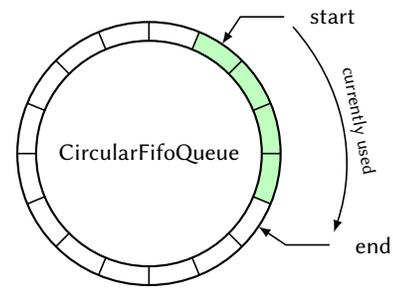
\begin{figure}
  \begin{tikzpicture}[>=latex,font=\sffamily,semithick,scale=1.75]
    \fill [green!25] (0,0) -- (67.5:1) arc [end angle=-22.5, start angle=67.5, radius=1] -- cycle;
    \draw [thick] (0,0) circle (1);
    \foreach \angle in {90,67.5,...,-67.5}
        \draw (\angle:1) -- (\angle-180:1);
    \node [circle,thick,fill=white,draw=black,align=center,minimum size=3cm] at (0,0) {CircularFifoQueue};
    \draw [<-] (56.25:1) -- (56.25:1.25) -- +(.333,0)
        node [right,inner xsep=.333cm] (Head) {start};
    \draw [<-] (-33.75:1) -- (-33.75:1.25) -- +(.333,0)
        node [right,inner xsep=.333cm] (Tail) {end};
    \draw [->,shorten >=5pt,shorten <=5pt] (Head.west) to [bend left] 
        node [midway,sloped,above,allow upside down] {\footnotesize currently used}
    (Tail.west);
  \end{tikzpicture}
  \caption{CircularFifoQueue Structure}
  \label{fig:circularFifoQueue}
\end{figure}

Using a \lstinline'CircularFifoQueue', the queue speed can be increased. The \lstinline'CircularFifoQueue' is a ring buffer for queue elements.  Its structure is visualized in Figure~\ref{fig:circularFifoQueue}: It contains an array of a given size, the index of the current start element and the index of the current end element. If a new element is added, the element at the end index is set and the end index is increased (modulo the queue size). If an element is taken from the queue, the element at the start index is returned, it is set to null and the start index is increased (modulo the queue size). Thereby, no new memory needs to be allocated while handling the queue. The current implementation of Apache Commons Collections does not react if the queue is full, i.e., if all elements are used, the new elements overwrite old elements.

Another problem of the basic \lstinline'CircularFifoQueue' is, that it is not directly usable in parallel. Since Kieker reading and writing is tone by different threads, this is necessary. Therefore, we use a tweaked version of the \lstinline'CircularFifoQueue' which is synchronized.\footnote{Synchronized version of \lstinline'CircularFifoQueue': \url{https://github.com/DaGeRe/KoPeMe/blob/main/kopeme-core/src/main/java/de/dagere/kopeme/collections/SynchronizedCircularFifoQueue.java}} To use it, we configured the queue using the Java properties that Kieker reads. 

\subsection{Storing Aggregated Data Instead of Method Executions}

Every insertion into the queue and every write to the hard disk is time-consuming. Additionally, the execution time depends on the current state of the hard disk. Therefore, we reduce the stored data by only obtaining and storing aggregated data, e.g., the average of 1000 method execution durations, instead of every method execution duration. This is achieved by adding two counters to each examined class, one which contains the sum execution time and one which contains the count of executions. For every iteration, the current execution time is added to the sum and the counter is increased by one. If the counter is equal to the parameterized execution count, for example 1000, a monitoring record is created. Additionally, the sum and the counter are reset.

This optimization makes it impossible to examine fine-grained properties of method invocations, e.g., the frequency of outliers. Therefore, it can only be applied if performance behaviour in the long run should be examined. For example, if a method became slower on 2\,\% on average, this could be examined using aggregated data storage. If outliers occur every 500 invocations, that cause an average slowdown of 0.5\,\%, this cannot be examined with aggregated data storage; in this case, the fine-grained data are required.


\section{Benchmarking Results of the Performance Optimizations}

To measure the overhead of our performance optimizations, we used the MooBench benchmark and executed the optimizations individually and the combination of them. In this section, we first describe our setup and afterwards the individual measurements.

\subsection{Setup}

For every optimization, we executed MooBench with and without the optimization. We used a call tree depth of 10, a method invocation count of 2\ 000\ 000 and a VM start count of 10, which are the default parameters of MooBench. Based on the technical suitability, we also compared the combination of different optimizations. Additionally, we compare the measurements to the execution without any instrumentation.

All measurements have been executed on an i7-4770 CPU @ 3.40GHz using the JVM from OpenJDK 1.8.0\_352, i.e., the latest OpenJDK release at the execution time. The following subsections describe our measurement results. All values are, if not otherwise specified, in \textbf{microseconds} ($\mu s$) per \textbf{MooBench method call}. For example, a measured value of 4.77 for Kieker with AspectJ and a call tree depth of 10 means that it took $4.77 \mu s$ for the monitoring of 10 nodes, i.e., a single node monitoring would have an average overhead of $0.477 \mu s$. The measurements can be repeated using the \lstinline'optimizations-all' branch of the MooBench main repository.\footnote{\url{https://github.com/kieker-monitoring/moobench/tree/optimizations-all}} The dataset of our measurement results is published.\footnote{\url{https://doi.org/10.5281/zenodo.7566677}} 

\subsection{Source Code Instrumentation}

The statistics of no instrumentation, AspectJ instrumentation and source code instrumentation are displayed in Table~\ref{tab:sci}. Due to the low mean execution time without instrumentation, the 95\,\% confidence interval has technically a size of 0. It shows that AspectJ instrumentation roughly increases the execution time by a factor of 10. Based on the granularity of methods, this factor might be higher or lower in production workloads. Using source code instrumentation roughly decreases the execution time by a factor of 2 and is therefore a recommendable optimization. However, it is only usable if the source code is available; therefore, using the \lstinline'-javaagent' to inject monitoring probes will in some cases stay necessary.

\begin{table}
  \begin{tabular}{crrr} \hline
 & No instr. & AspectJ & Source Code Instr.\\ \hline
Mean & 0.0548 & 4.7711 & 2.4169\\
95\,\% & $\pm$ 0.0 & $\pm$ 0.0444 & $\pm$ 0.0038 \\ \hline
$Q_1$ & 0.0520 & 4.1890 & 2.4380\\
Median & 0.0530 & 5.0790 & 2.5730\\
$Q_3$ & 0.0580 & 5.5030 & 2.6880\\
\hline
  \end{tabular}
  \caption{Statistics of Source Code Instrumentation}
  \label{tab:sci}
\end{table}

\subsection{DurationRecord}
\label{kap:durationRecord}

Since the \lstinline'DurationRecord' is currently only implemented for the source code instrumentation, we compared the source code instrumentation for Kiekers regular \lstinline'OperationExecutionRecord' and the source code instrumentation using the \lstinline'DurationRecord'. The results are displayed in Table~\ref{tab:durationRecord}. 
It shows that using \lstinline'DurationRecord' also reduces the measurement duration statistically significant, but not with an effect size in the order of magnitude of the improvement using source code instrumentation. 

\begin{table}
  \begin{tabular}{crrr} \hline
 & No instr. & Source Code Instr. & DurationRecord\\ \hline
Mean & 0.0548 & 2.4169 & 2.3426 \\
95\,\% & $\pm$ 0.0 & $\pm$ 0.0038 & $\pm$ 0.0037 \\ \hline
$Q_1$ & 0.0520 & 2.4380 & 2.1720\\
Median & 0.0530 & 2.5730 & 2.4720\\
$Q_3$ & 0.0580 & 2.6880 & 2.6080\\
\hline
  \end{tabular}
  \caption{Statistics of \lstinline'DurationRecord'}
  \label{tab:durationRecord}
\end{table}

\subsection{CircularFifoQueue}

The \lstinline'CircularFifoQueue' can be used with source instrumentation and either \lstinline'OperationExecutionRecord' or \lstinline'DurationRecord'. Therefore, we compare the execution times of both. The results are displayed in Table~\ref{tab:durationRecord}. 
The table shows that the \lstinline'CircularFifoQueue' provides a statistically significant performance improvement, both when using \lstinline'OperationExecutionRecord' and \lstinline'DurationRecord'. Surprisingly, the improvement is higher when using Kiekers default  \lstinline'OperationExecutionRecord' than when using \lstinline'DurationRecord'. We assume that this is due to internal optimizations of the JVM; however, combining \lstinline'CircularFifoQueue' and \lstinline'DurationRecord' should therefore not be blindly applied.

\begin{table}[h]
  \begin{tabular}{crrr} \hline
 & No instr. & CircularFifoQueue & CircularFifoQueue\\ 
 & & + OperationExecutionR. & + DurationRecord\\ \hline
Mean & 0.0548 & 1.5017 & 1.6503\\
95\,\% & $\pm$ 0.0 & $\pm$ 0.0047 & $\pm$ 0.0039 \\ \hline
$Q_1$ & 0.0520 & 1.2330 & 1.2670\\
Median & 0.0530 & 1.3300 & 1.4190\\
$Q_3$ & 0.0580 & 1.4810 & 1.7420\\
\hline
  \end{tabular}
  \caption{Statistics of \lstinline'CircularFifoQueue'}
  \label{tab:durationRecord}
\end{table}

Additionally, the \lstinline'CircularFifoQueue' could \textit{swallow} elements if more elements than the capacity of the queue are added. Therefore, we would only advise using \lstinline'CircularFifoQueue' in specific settings, where it can be guaranteed that this does not happen. Another method for reduction of the queue overhead is the aggregation of data before they are inserted into the queue, is presented in the following.

\subsection{Aggregated Writing}

Aggregating is currently only implemented for source code instrumentation. Additionally, it is only implementable straightforward for \lstinline'DurationRecord', since metadata like the execution stack size might not be aggregated easily. Therefore, we compare the aggregated writing and usage of \lstinline'DurationRecord'. There might be mechanisms for aggregating also metadata, e.g., by storing a mapping from the execution stack size to the current data. The performance characteristics of this would heavily depend on the tree structure. Examining this could be a part of future work.

Table~\ref{tab:aggregated} shows the statistics of aggregated data reading (when aggregating always 1000 method invocations). It shows that aggregating the data also significantly decreases the monitoring overhead. Therefore, for use cases where aggregated data can be used, we would advise using aggregated writing.

\begin{table}[h]
  \begin{tabular}{crrr} \hline
 & No instr. & DurationRecord & Aggregated Writing\\ \hline
Mean & 0.0548 & 2.3426 & 0.4014\\
95\,\% & $\pm$ 0.0 & $\pm$ 0.0037 &$\pm$ 0.0003\\ \hline
$Q_1$ & 0.0520 & 2.1720 & 0.3810\\
Median & 0.0530 & 2.4720 & 0.3860 \\
$Q_3$ & 0.0580 & 2.6080 & 0.3870\\
\hline
  \end{tabular}
  \caption{Statistics of aggregated data reading}
  \label{tab:aggregated}
\end{table}

\subsection{Combination}

Since aggregated writing reduces the amount of created monitoring records, it also makes sense to combine aggregated writing and the usage of the \lstinline'CircularFifoQueue'. Table~\ref{tab:all} shows the statistics of this approach. In this setting, the difference in the measured values is not statistically significant. Therefore, we cannot make a statement about whether using the \lstinline'CircularFifoQueue' leads to an overhead reduction in this setting.

\begin{table}[h]
  \begin{tabular}{crrr} \hline
 & No instr. & Aggregated Writing & Combination \\ \hline
Mean & 0.0548 & 0.4014 & 0.3897\\
95\,\% & $\pm$ 0.0 & $\pm$ 0.0003 & $\pm$ 0.0002 \\ \hline
$Q_1$ & 0.0520 & 0.3810 & 0.3830 \\
Median & 0.0530 & 0.3860 & 0.3880 \\
$Q_3$ & 0.0580 & 0.3870 & 0.3890 \\
\hline
  \end{tabular}
  \caption{Statistics of aggregated data reading}
  \label{tab:all}
\end{table}

\begin{figure*}
  \includegraphics[width=16cm]{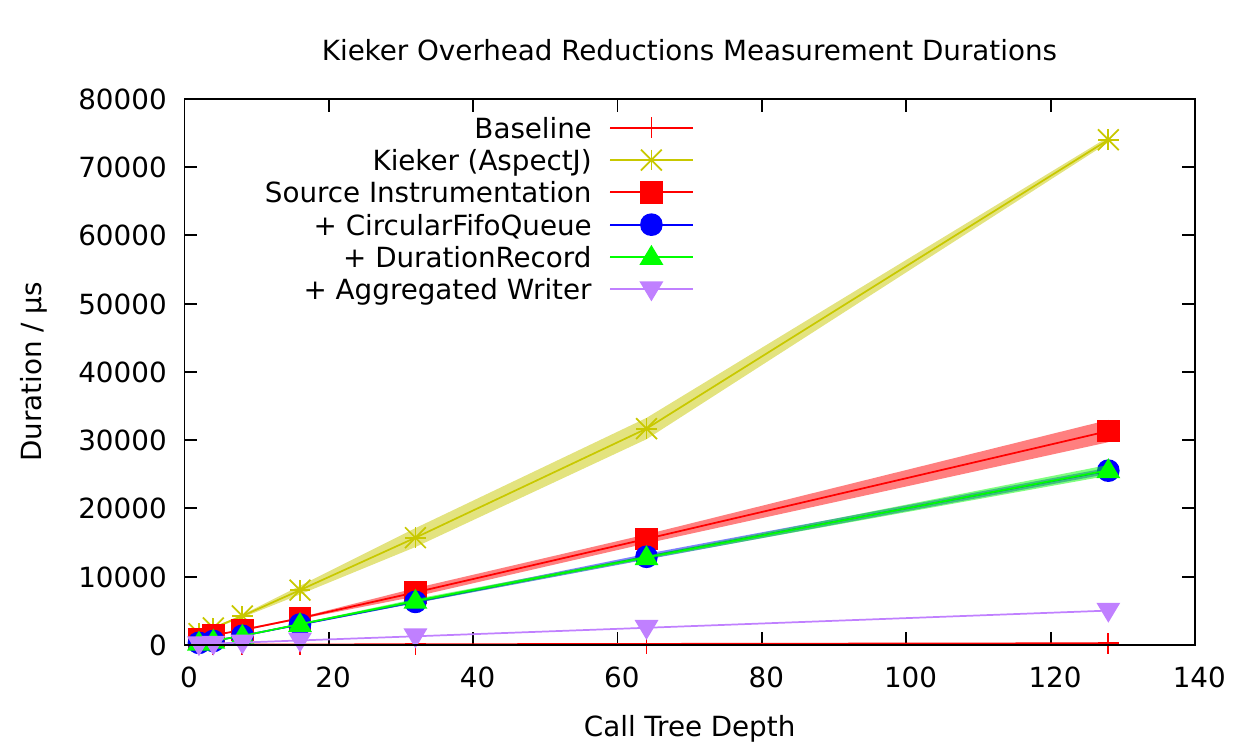}
  \caption{Growth of the Overhead Based on Call Tree Size}
  \label{fig:callTree}
\end{figure*}

In real-world use cases, the overhead is not only relevant for a call tree depth of 10. Therefore, we measured how the combination of these optimizations perform for different call tree sizes. Figure~\ref{fig:callTree} shows how the overhead evolves with growing call tree depth. For better visibility, the measurement points are connected, even if the call tree depth is a concrete value. The areas around the curves marks the area of $\mu + \sigma$ and $\mu - sigma$, where $\mu$ is the mean and $\sigma$ is the standard deviation.
We see adding the use of \lstinline'DurationRecord' only slightly reduces the overhead for the examined call tree sizes. Since using the \lstinline'DurationRecord' also reduces the overhead if we do not use the \lstinline'CirclarFifoQueue', and since the metadata are not usable when aggregating operation calls, we keep using the \lstinline'DurationRecord'. Overall, the figure shows that for a call tree size of 128, all optimizations reduce the overhead significantly in comparison to the regular monitoring using AspectJ.

\section{Related Work}

Related work can be grouped into work that only \textit{measures the monitoring overhead}, work that examines ways to \textit{reduce the monitoring overhead} and \textit{work that reduces the monitoring} itself. In the following, we give an overview over works from these fields.

\subsection{Measurement}

Horky et al. \cite{horkyOverhead} examine the overhead of dynamic monitoring in Java, i.e., monitoring that is inserted and removed by demand. They state that three effects--the mere presence of the probe, the manipulation of the probes execution and the code optimizations possibly happening because of the probe--influence the execution time of the monitored method. By using the SPECjbb2015\texttrademark, they find that the activation and deactivation of probes in realistic environments might both slow down or speed up the execution. 

Bara et al. \cite{bara2015hardware} develop hardware support for performance monitoring in an processor core and show that performance monitoring also increases the energy consumption of a processor. In contrast to this work, they focus on hardware level performance monitors.

A main requirement for performance measurement is reproducability. Eichelberger et al. \cite{eichelberger2016reproducibility} research how MooBench measurements of SPASS-meter can be reproduced. Additionally, Knoche and Eichelberger \cite{knoche2017raspberry,knoche2018using} discuss how a Raspberry Pi can be used for reproducable performance measurement. In contrast to their work, we present measurement results of an established benchmark on a typical Desktop PC. In order to facilitate reproducibility of our result, it would make sense to repeat our measurements on different execution hardware, e.g., on a Raspberry Pi.


Furtheremore, there exist measurement implementations that aim for minimal overhead using different techniques. JPortal \cite{zuo2021jportal} achieves minimal overhead by using hardware-based tracing using Intel processor trace. Since this requires decoding the stored metadata of maschine code executions, the obtained traces are not completely correct. However, they report an overhead of at maximum 16.5\,\%. In contrast to our work, they used different workloads for benchmarking and their goal was not to get measure accurate runtimes but obtain accurate traces.

Schardl et al. \cite{schardl2017csi} develop the CSI framework, which is able to insert code at compile time. This can--amongst other use cases--be utilized to measure the performance of executed code. They evaluate the CSI framework by instrumenting the Apache Server and the bzip2 data compressor and benchmarking them by appropriate benchmarks. They find that the overhead is less than 70\,\% of the programs execution time. While they pursue a similiar approach for instrumentation, e.g., instrumenting at compile time, they use a different technology (software written in c) and evaluate benchmarks measuring real-world use cases. Therefore, their results are not directly compareable to ours.

\subsection{Reduction of the Monitoring Overhead}

In the following, we first describe the research on overhead reduction in Kieker chronologically and afterwards the work on using a different instrumentation technique.

To \textit{reduce the the monitoring overhead}, Waller and Hasselbring \cite{waller2012comparison} first researched how the use of multi-core processors influence the runtime overhead of Kieker. They find that asynchronous monitoring writers have lower overhead when multi-core processor systems are used.
Waller et al. \cite{waller2014application} state that the monitoring overhead consists of the instrumentation overhead, the data collection time and the time for writing the data to a queue. They examine four possible optimizations:
\begin{inparaenum}
  \item Internal optimizations, e.g., reusing the method \lstinline'String' definition,
  \item using an \lstinline'ArrayBlockingQueue' from the disruptor framework,
  \item sending \lstinline'ByteBuffer' instances to the monitoring queue, instead of the record, and
  \item reducing the amount of source used in Kieker by providing a minimal Kieker project.
\end{inparaenum}
They find that the first two optimizations provide significant improvements and are feasible from a maintainability point of view. Optimization (3) reduces the overhead, but requires the definition of individual serializations for each record type, and is therefore not considered further. Optimization (4) does slightly reduce the overhead, but also reduces the maintainability and extendability of Kieker and is therefore also not considered further. 

Finally, Strubel and Wulf \cite{strubel2016refactoring} further reduce the monitoring overhead. Formerly, string attributes were serialized using a registry records having an id, so strings are not fully serialized. This transformation of registry records was done in the application thread; by moving it to the writer thread, both the code complexity and the overhead inside of the application thread could be reduced.

Since this work builds on the current Kieker version, the feasible optimizations created by Waller and Hasselbring \cite{waller2012comparison}, Waller et al. \cite{waller2014application} and Strubel and Wulf \cite{strubel2016refactoring} are already included. Nevertheless, they had the focus of preserving the architecture discovery functionality of focus, whereas this work focusses on monitoring method execution duration without preserving the architecture discovery functionality.


The domain-specific language for bytecode instrumentation (DiSL) enables a more fine-grained instrumentation than AspectJ \cite{marek2012disl}. Since it relies on ASM, it is very likely to have different performance characteristics than Kieker's default AspectJ instrumentation. Therefore, it would be an alternative for our source code instrumentation and could therefore be evaluated against it. However, there is currently no usable implementation provided, therefore we cannot compare them.

Okanovic et al. \cite{okanovic2013dprof} examined how to \textit{use different instrumentation techniques} in order to reduce the monitoring overhead. Therefore, they used DiSL instead of AspectJ for their monitoring system DProf. They find that using DiSLs reduced the monitoring overhead by 1.2\,\% for their use case. In contrast to our work, they used their system DProf to evaluate the performance overhead. Their performance improvement by 1.2\,\% indicates that DiSL very likely does not offer a overhead reduction that is as big as the improvement of our source instrumentation. 

\subsection{Reduction of the Monitoring}

For reduction of monitoring itself, Popiolek et al. \cite{popiolek2021lowoverhead} research how to reduce the overhead of performance monitoring by reducing the performance monitors in cloud infrastructures. They do so by clustering the counters using the Pearson correlation coefficient, i.e., by determining which counters are either high or low at the same time. Thereby, they reduce the monitoring overhead by up to one third. Similarly, Shang et al. \cite{shang2015automated} determine the performance counters also by using Pearson correlation of the measured values. Afterwards, they hierarchically cluster the performance counters using Calinski-Harabasz stopping rule. In contrast to these works, we do not try to select where to monitor but to reduce the monitoring overhead at the methods where monitoring happens.

Mertz and Nunes \cite{mertz2019onthefeasibility} reduce the overhead by adaptive configuration: In the first step, they perform a lightweight monitoring. Based on the results of this monitoring, they decide which parts of the source code should be monitored in more detail. In contrast to this work, we do not perform an adaptive configuration but try to reduce monitoring overhead with static configuration.

\section{Summary and Outlook}

We described how we reduced the monitoring overhead of the Kieker framework. This was done using four changes to the monitoring process: Instrumenting the source code directly instead of instrumenting using AspectJ, reducing the stored monitoring data, using \lstinline'CircularFifoQueue' instead \lstinline'LinkedBlockingQueue' and using an aggregated writer instead of Kiekers \lstinline'FileWriter'. 

Reduction of the stored monitoring data and usage of an aggregated writer is only feasible if the long-term performance should be examined. If individual method invocations and outliers for these invocations should be examined, only source code instrumentation and the queue exchange can be used.

Using the monitoring overhead benchmark MooBench, we showed that all of these optimizations decrease the monitoring overhead. Therefore, it is planned to make them available for regular monitoring. This requires the following adaptions:
\begin{inparaenum}[(1)]
  \item For Kieker source instrumentation, this has been already done by publishing the code as part of the Kieker GitHub organization. 
  \item For the \lstinline'DurationRecord', the process of including records with reduced monitoring information into the Kieker infrastructure is currently ongoing. Since Kieker supports a variety of languages, the inclusion of such records requires defining a basic record in the Kieker Instrumentation Language and generating the records for every language out of it.
  \item Switching the queue is a functionality that will stay up to the user. While using the \lstinline'CircularFifoQueue' improves the performance statistically significant, it is unclear whether production systems might produce too many records for the queue and therefore records get lost. Future research might examine whether adding additional checks for integrity could be added without disturbing the monitoring overhead reduction.
  \item It is also planned to add the \lstinline'AggregatedWriter' as part of Kieker itself, but not set it as default. Thereby, users can switch to the \lstinline'AggregatedWriter' if they use the \lstinline'DurationRecord' and aim for only measuring the performance, but not doing architecture recovery tasks.
\end{inparaenum}
In addition to adapting Kieker itself, it should also be checked to which degree aggregating the data improves the measurement quality. While we assume that lower overhead always results in improved distinguishability of measurement results, this thesis requires checking.

This work is a first step towards minimal overhead monitoring: In addition to optimize the process itself, the following further optimizations should be researched:
\begin{inparaenum}[(1)]
  \item Our optimized monitoring process should be compared to sampling. While sampling is only able to obtain the stack trace at safepoints, it has a very low overhead. Therefore, a comparison between our optimized Kieker probe and sampling should compare both, the overhead and the ability to detect performance changes. 
  \item Usage of adaptive monitoring (like \cite{mertz2019onthefeasibility}): It was shown that deactivated probes nearly reduce the monitoring overhead to 0. To obtain usable data while preserving measurement accuracy, it might be promising to activate probes only part-time. Additionally, our approach is only able to measure the performance and not the creation of the call tree, since no monitoring metadata are present. If monitoring metadata are required, it would also be possible to add multiple probes to the same source code, one regular probe for creation of \lstinline'OperationExecutionRecord' and one probe for \lstinline'DurationRecord'. Afterwards, the necessary probes could be used.
\end{inparaenum}
By this work and the examination of sampling and adaptive monitoring, the definition of minimal overhead processes for performance monitoring can be created. This could be the basis for understanding the performance behavior of software on a low level in production systems.

\textbf{Acknowledgments} This work was funded by the German Federal Ministry of Education and Research within the project “Performance Überwachung Effizient Integriert” (PermanEnt, BMBF 01IS20032D).




\end{document}